\documentclass[nofootinbib,showpacs,aps,prl,twocolumn,superscriptaddress,floatfix]{revtex4}

\usepackage{graphicx} 

\begin{document}

\newcommand{\ee}{{$e^{+}e^{-}$}}
\newcommand{\mev}{MeV/$c^2$}
\newcommand{\gev}{GeV/$c^2$}
\newcommand{\phiee}{$\phi \rightarrow e^{+}e^{-}$}
\newcommand{\nphi}{$N_{\phi}$}
\newcommand{\nex}{$N_{\rm ex}$}
\newcommand{\ephir}{$N_{\rm ex}/(N_{\phi}+N_{\rm ex})$}
\newcommand{\bg}{$\beta \gamma$}
\newcommand{\omegaee}{$\omega \rightarrow e^{+}e^{-}$}
\newcommand{\lambdappi}{$\Lambda \rightarrow p \pi^-$}
\newcommand{\kspipi}{$K^{0}_s \rightarrow \pi^+\pi^-$}

\title{Evidence for in-medium modification of $\phi$ meson at normal nuclear density}

\author{R.~Muto}
\email{muto@riken.jp}
\affiliation{RIKEN, 2-1 Hirosawa, Wako, Saitama 351-0198, Japan}
\author{J.~Chiba}
\altaffiliation[Present Address: ]{Faculty of Science and Technology, Tokyo University of Science, 2641 Yamazaki, Noda, Chiba 278-8510, Japan}
\affiliation{Institute of Particle and Nuclear Studies, KEK, 1-1 Oho, Tsukuba, Ibaraki 305-0801, Japan}
\author{H.~En'yo}
\affiliation{RIKEN, 2-1 Hirosawa, Wako, Saitama 351-0198, Japan}
\author{Y.~Fukao}
\affiliation{Department of Physics, Kyoto University, Kitashirakawa, Sakyo-ku, Kyoto 606-8502, Japan}
\author{H.~Funahashi}
\affiliation{Department of Physics, Kyoto University, Kitashirakawa, Sakyo-ku, Kyoto 606-8502, Japan}
\author{H.~Hamagaki}
\affiliation{Center for Nuclear Study, Graduate School of Science,\\
University of Tokyo, 7-3-1 Hongo, Tokyo 113-0033, Japan}
\author{M.~Ieiri}
\affiliation{Institute of Particle and Nuclear Studies, KEK, 1-1 Oho, Tsukuba, Ibaraki 305-0801, Japan}
\author{M.~Ishino}
\altaffiliation[Present Address: ]{ICEPP, University of Tokyo, 7-3-1 Hongo, Tokyo 113-0033, Japan}
\affiliation{Department of Physics, Kyoto University, Kitashirakawa, Sakyo-ku, Kyoto 606-8502, Japan}
\author{H.~Kanda}
\altaffiliation[Present Address: ]{Physics Department, Graduate School of Science, Tohoku University, Sendai 980-8578, Japan}
\affiliation{Department of Physics, Kyoto University, Kitashirakawa, Sakyo-ku, Kyoto 606-8502, Japan}
\author{M.~Kitaguchi}
\affiliation{Department of Physics, Kyoto University, Kitashirakawa, Sakyo-ku, Kyoto 606-8502, Japan}
\author{S.~Mihara}
\altaffiliation[Present Address: ]{ICEPP, University of Tokyo, 7-3-1 Hongo, Tokyo 113-0033, Japan}
\affiliation{Department of Physics, Kyoto University, Kitashirakawa, Sakyo-ku, Kyoto 606-8502, Japan}
\author{K.~Miwa}
\affiliation{Department of Physics, Kyoto University, Kitashirakawa, Sakyo-ku, Kyoto 606-8502, Japan}
\author{T.~Miyashita}
\affiliation{Department of Physics, Kyoto University, Kitashirakawa, Sakyo-ku, Kyoto 606-8502, Japan}
\author{T.~Murakami}
\affiliation{Department of Physics, Kyoto University, Kitashirakawa, Sakyo-ku, Kyoto 606-8502, Japan}
\author{T.~Nakura}
\affiliation{Department of Physics, Kyoto University, Kitashirakawa, Sakyo-ku, Kyoto 606-8502, Japan}
\author{M.~Naruki}
\affiliation{RIKEN, 2-1 Hirosawa, Wako, Saitama 351-0198, Japan}
\author{K.~Ozawa}
\altaffiliation[Present Address: ]{Department of Physics, University of Tokyo, 7-3-1 Hongo, Tokyo 113-0033, Japan}
\affiliation{Center for Nuclear Study, Graduate School of Science,\\
University of Tokyo, 7-3-1 Hongo, Tokyo 113-0033, Japan}
\author{F.~Sakuma}
\affiliation{Department of Physics, Kyoto University, Kitashirakawa, Sakyo-ku, Kyoto 606-8502, Japan}
\author{O.~Sasaki}
\affiliation{Institute of Particle and Nuclear Studies, KEK, 1-1 Oho, Tsukuba, Ibaraki 305-0801, Japan}
\author{M.~Sekimoto}
\affiliation{Institute of Particle and Nuclear Studies, KEK, 1-1 Oho, Tsukuba, Ibaraki 305-0801, Japan}
\author{T.~Tabaru}
\affiliation{RIKEN, 2-1 Hirosawa, Wako, Saitama 351-0198, Japan}
\author{K.~H.~Tanaka}
\affiliation{Institute of Particle and Nuclear Studies, KEK, 1-1 Oho, Tsukuba, Ibaraki 305-0801, Japan}
\author{M.~Togawa}
\affiliation{Department of Physics, Kyoto University, Kitashirakawa, Sakyo-ku, Kyoto 606-8502, Japan}
\author{S.~Yamada}
\affiliation{Department of Physics, Kyoto University, Kitashirakawa, Sakyo-ku, Kyoto 606-8502, Japan}
\author{S.~Yokkaichi}
\affiliation{RIKEN, 2-1 Hirosawa, Wako, Saitama 351-0198, Japan}
\author{Y.~Yoshimura}
\affiliation{Department of Physics, Kyoto University, Kitashirakawa, Sakyo-ku, Kyoto 606-8502, Japan}
\collaboration{KEK-PS E325 Collaboration}

\date{\today}

\begin{abstract}
Invariant mass spectra of \ee pairs have been measured in 12 GeV $p+A$ reactions
to detect possible in-medium modification of vector mesons.
Copper and carbon targets are used to study the nuclear-size dependence
of \ee\ invariant mass distributions.
A significant excess on the low-mass side of the $\phi$ meson peak is
observed in the low $\beta\gamma ( = \beta/\sqrt{1-\beta^2} )$ region of $\phi$ mesons ( $\beta\gamma < 1.25$ ) with copper targets.
However, in the high $\beta\gamma$ region
$(\beta\gamma > 1.25)$, spectral shapes of $\phi$ mesons are well
described by the Breit-Wigner shape when experimental effects are considered.
Thus, in addition to our earlier publications on $\rho / \omega$
modification, this study has experimentally verified
vector meson mass modification at normal nuclear density.
\end{abstract}


\maketitle

The properties of hadrons, which are composite particles of quarks and gluons,
have been measured and determined in the past in many experimental studies.
These properties, such as mass and decay width,
represent vacuum expectation values. Therefore, in principle,
they can differ when the vacuum itself changes, i.e.,
when they are in hot and/or dense matter.

A possible and interesting experimental approach to the in-medium
properties of hadrons is to measure the di-leptons from the decays of
hadrons
in hot/dense nuclear matter, in which the chiral symmetry
could be restored
leading to changes in those properties.
The present experiment, KEK-PS E325, was conducted at the KEK 12-GeV
Proton-Synchrotron, in order to search for in-medium mass modifications
of vector mesons in the reaction 12~GeV $p + A \to \rho, \omega, \phi + X \to e^+e^- + X'$.
In our earlier publications~\cite{ozawa,naruki},
we reported the mass modification of $\rho$ and $\omega$ mesons in a nuclear
medium.
In the present paper, we report new results of the $\phi$ meson.

In the case of the $\rho$ and $\omega$ mesons, it is very difficult to 
distinguish between matter effects
affecting only one of the two mesons, or both,
due to the overlapping of two peaks with different widths~\cite{ozawa,naruki}.
The natural width of the $\phi$ meson is narrow (4.26 \mev)
without other resonances in the vicinity;
therefore, we can examine the possible mass modification more clearly.
When the properties are modified by the medium,
the observed spectrum contains two components:
in-vacuum decays with a normal mass distribution and
in-medium decays with a modified mass distribution.
The latter component can cause an excess around the $\phi$ meson peak
in the invariant mass spectra.

Theoretically, the possibility of the decrease in the mass of
light vector mesons in
a nuclear medium was first pointed out by Brown and Rho~\cite{br}.
Thereafter, many theoretical studies were conducted.
Hatsuda and Lee calculated the density dependence of the $\phi$ meson mass 
based on QCD sum rules~\cite{HL92}.
According to their calculation,
the expected mass decrease for the $\phi$ meson at normal nuclear density is
20--40 \mev .
As for the decay width of the $\phi$ meson,
some theoretical calculations predict the broadening of
the width by a factor between five or six~\cite{oset,cabrera} and ten~\cite{KWW}, at normal nuclear density.
When the width of the $\phi$ meson broadens by a factor of ten,
the lifetime of a $\phi$ meson, $c\tau_{\phi}$, 
in a nucleus is reduced from $46$ to $5\ {\rm fm}$ and
the probability of in-medium decay increases, thereby reflecting the in-medium
properties.

Although several experimental reports
on the in-medium modification of $\rho$ and $\omega$
mesons exist, including our reports~\cite{ceres,tagx,taps,na60,ozawa,naruki}
\footnote{Recently, the GSI S236 collaboration reported an enhancement of the pion-nucleus potential in a deeply bound pionic $^{115,119,123}\rm{Sn}$ nuclei as a possible signature of chiral symmetry restoration~\cite{pionic}.}
, experimental information on the $\phi$ meson modification is very limited.
The CERES/NA45 experiment measured dielectron spectra in high-energy heavy-ion
collisions~\cite{ceres}; the HADES experiment measured
dielectron spectra in 2 GeV$\cdot$A C-C collisions~\cite{hades}.
However, in both experiments, the mass resolution and/or statistics
for the $\phi$ mesons were not sufficient
to draw a definite conclusion regarding the $\phi$ meson mass modification.
The NA60 collaboration measured the dimuon spectra
in 158 GeV$\cdot$A In-In collisions~\cite{na60},
and the PHENIX experiment at RHIC reported the \phiee\ spectrum
in Au+Au collisions at $\sqrt{s_{NN}}=200\ {\rm GeV}$~\cite{ozawaphenix}.
Recently, the LEPS collaboration reported a possible $\sigma_{\phi N}$ modification
in medium by measuring the $A$-dependence of the $\phi$
photo-production yields in the $K^+K^-$ decay mode~\cite{ishikawa}.
Thus far, no clear evidence for the modification of
the $\phi$ meson mass has been observed in the above experiments.
The result described in the present paper is the first positive signal
of the $\phi$ meson modification.

Detector elements relevant to our analysis are briefly described as follows.
For further details of the E325 spectrometer, see \cite{nim}.
It comprises two arms with electron ID counters and kaon ID counters
that share a dipole magnet and tracking devices.
The typical acceptance in the laboratory frame was $0.5<{\rm rapidity}<2.0$ and $1 < \beta\gamma<3 $ for \ee\ pairs. 
In the present paper, we report analysis results with \ee-triggered data
collected in 2001 and 2002.
A primary proton beam with a typical intensity of 9 (7) $\times$ 10$^8$ per 1.8-sec spill
in 2001 (2002) was
delivered to targets located at the center of the magnet.
In order to observe the nucleus-size dependence, we accumulated data
by using two types of targets, carbon and copper.
In 2001, one carbon and two copper targets were used simultaneously,
while in 2002,
one carbon and four copper targets were used simultaneously.
The thickness of each copper target was
73 mg/cm$^{2}$ and that of the carbon target was
92 (184) mg/cm$^{2}$ in 2001 (2002).
They were aligned along the beam axis and
separated typically by 46 (23) mm in 2001 (2002).

In order to reproduce the observed invariant mass spectra,
we performed a detailed detector simulation using {\scshape Geant4}~\cite{g4}.
All the experimental effects that affect the invariant mass spectrum,
such as
multiple scattering and energy loss including the external Bremsstrahlung of particles,
tracking performance with chamber resolution,
and misalignment of tracking devices, were considered.
The effect of internal radiative corrections was also taken into account
according to~\cite{ib}. 
The mass resolution of \phiee\ was estimated to be 10.7 \mev.

\begin{figure}
\includegraphics[width=8.6cm,height=10.7cm]{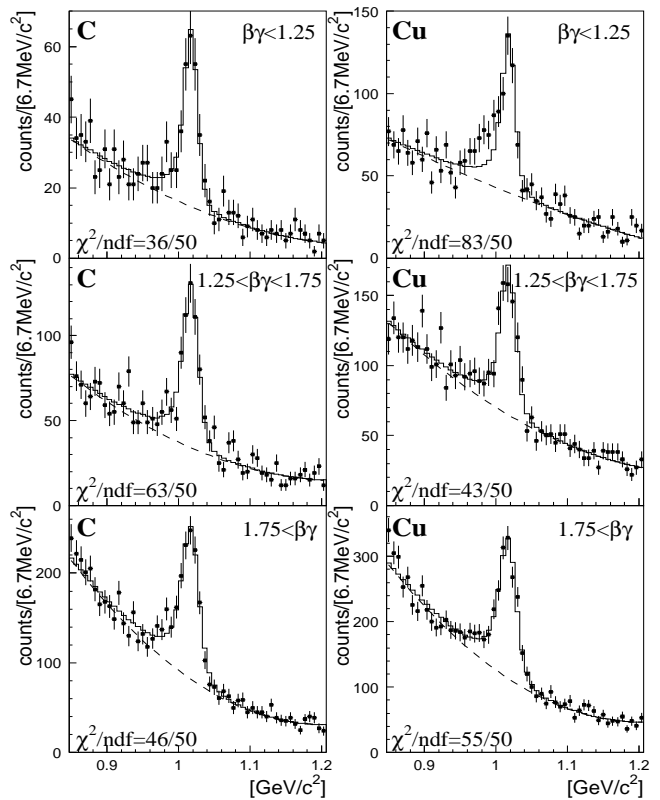}
\caption{Obtained $e^+e^-$ distributions with the fit results.
The target and $\beta\gamma$-region are shown in each panel.
The points with error bars represent the data.
The solid lines represent the fit results with an expected \phiee\ shape
and a quadratic background.
The dashed lines represent the background.
\label{fig:invmassall}}
\end{figure}

\begin{figure}
\includegraphics[width=5.9cm,height=2.5cm]{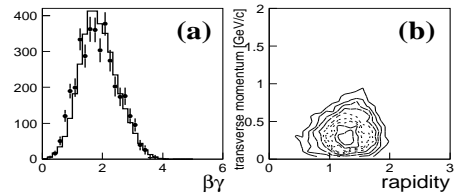}
\caption{Kinematical distributions of \ee\ pairs in the mass region $0.95 < M_{ee} < 1.05\ {\rm GeV}/c^2$. Acceptance was not corrected.
Distributions of (a) $\beta\gamma$ and (b) contours in transverse momentum and rapidity.
In the plot (a), points represent data and the line represents simulation result
using the nuclear cascade code, {\scshape JAM}~\cite{jam}.
\label{fig:kin}}
\end{figure}

We reconstructed the masses of the $\phi$ mesons from the measured
momenta of the $e^+$ and $e^-$.
Figure~\ref{fig:invmassall} shows the obtained invariant mass distributions.
We divided the data into three parts based on the $\beta \gamma$ values
of the observed \ee\ pairs,
$\beta \gamma < 1.25$, $1.25 < \beta\gamma < 1.75$, and $1.75 < \beta\gamma$.
We fitted each mass spectrum with a resonance shape of \phiee\ and
a quadratic background curve.
For the $\phi$ meson resonance shape, we used the Breit-Wigner curve
$M_\phi(m) \propto 1/((m-m_0)^2+(\Gamma_0/2)^2)$ 
with pole mass $m_0 = 1019.456$\ \mev\ and decay width
$\Gamma_0 = 4.26$\ \mev\ convoluted over the detector response
in the simulation
according to the kinematical distributions of the $\phi$ mesons
in each $\beta\gamma$ region.
The kinematical distributions of the $\phi$ meson were obtained by the nuclear cascade
code {\scshape JAM}~\cite{jam},
which reproduced well the observed distributions as shown in Fig.~\ref{fig:kin}.
The relative abundance of the $\phi$ mesons \nphi,
and the parameters of the quadratic
background were obtained from the fit.
The fit region was from $0.85$ to $1.2\ {\rm GeV}/c^2$.
The carbon data were well reproduced by the fit
in all the $\beta \gamma$ regions.
On the other hand,
the copper data in the lowest $\beta\gamma$ region contradicted the
applied resonance shape at $99\%$ C.L.
due to the visible excess on the low mass side of the $\phi$ meson peak.

To evaluate the amount of excess \nex,
we repeated the same fit procedure
except that the mass region from $0.95$ to $1.01\ {\rm GeV}/c^2$,
where the excess is observed,
was excluded from the fit.
The obtained $\chi^2/ndf$'s including and excluding the excess region
are shown in TABLE~\ref{tab:chi2}.
We obtained \nex\ by subtracting the integral of the fit results from
that of the data
in the mass region from $0.95$ to $1.01\ {\rm GeV}/c^2$.
To evaluate systematic errors,
we varied the fit region, binning, mass resolution and mass scale,
changed the background curve from quadratic to cubic,
changed the Breit-Wigner shape
from the non-relativistic one to the relativistic one,
and modified the kinematical distribution of $\phi$
mesons in the simulation.
Then we averaged the obtained values of \nex\ and
\nphi\ to obtain the listed values in TABLE~\ref{tab:ne}.
The systematic errors represent the maximum deviations
from the averaged values.
For the statistical errors we selected the largest values
in above fit conditions.
The excess is statistically significant
for the copper target data in the lowest $\beta\gamma$ bin,
whereas it is marginal for the carbon target data.
This excess is considered to be the signal of
the mass modification of the $\phi$ mesons in a target nucleus
because such an effect should be visible only for slow
$\phi$ mesons produced in a large target nucleus.

In the mass region of the $\rho$ meson
the excess was observed both in C and Cu targets~\cite{naruki},
but in the case of the $\phi$ meson we observed the excess only in Cu targets data
with small $\beta \gamma$.
The $\rho$-meson lifetime ($c\tau \sim 1.3\ {\rm fm}$) is shorter than
the typical nuclear size and a significant portion of $\rho$ mesons
should decay in both C and Cu nuclei.
However, the $\phi$-meson lifetime ($c\tau \sim 46\ {\rm fm}$) is much longer,
so even if it is modified in nuclear medium,
such an effect should only be visible for a larger nucleus and for
$\phi$'s with small $\beta \gamma$.

\begin{table}
\caption{
The fit $\chi^2/ndf$ values (a) including
and (b) excluding the excess region: $0.95-1.01\ {\rm GeV}/c^2$.
\label{tab:chi2}}
\begin{tabular}{c|c|cc}
&$\beta\gamma$ range&$\chi^2/ndf$ (a)& $\chi^2/ndf$ (b)\\\hline
&$<1.25$ &$36/50$ & $31/41$\\
C&$1.25-1.75$ & $63/50$ & $54/41$\\
&$1.75<$ &$46/50$&$37/41$\\\hline\hline
&$<1.25$ &$83/50$ & $57/41$\\
Cu&$1.25-1.75$ &$43/50$&$38/41$ \\
&$1.75<$ & $55/50$&$53/41$ \\
\end{tabular}
\end{table}
\begin{table}
\caption{Numbers of $\phi$ mesons $[N_{\phi}]$ and excess $[N_{ex}]$ evaluated 
by the fits (see text). Error values shown are statistic (first) and systematic (second).
\label{tab:ne}}
\begin{ruledtabular}
\begin{tabular}{c|c|ccc}
&$\beta\gamma$ range&\nex & \nphi & \nex/(\nex+\nphi)\\\hline
&$<1.25$ & $6\pm17^{+7}_{-6}$& $257\pm26^{+6}_{-7}$ & $0.02\pm0.06^{+0.02}_{-0.01}$\\
C&$1.25 - 1.75$ & $-4\pm26^{+10}_{-12}$ & $547\pm41^{+13}_{-15}$ & $-0.01\pm0.05^{+0.02}_{-0.01}$\\
&$1.75<$ & $39\pm42^{+22}_{-25}$ & $1076\pm64^{+12}_{-15}$ &$0.04\pm0.04^{+0.02}_{-0.01}$\\\hline\hline
&$<1.25$ & $133\pm28^{+5}_{-12}$ & $464\pm38^{+6}_{-5}$ &$0.22\pm0.04^{+0.01}_{-0.01}$\\
Cu&$1.25-1.75$ & $35\pm33^{+9}_{-12}$ & $588\pm47^{+14}_{-8}$ &$0.06\pm0.05^{+0.01}_{-0.01}$ \\
&$1.75<$ & $21\pm48^{+25}_{-29}$ & $1367\pm72^{+24}_{-27}$ &$0.02\pm0.03^{+0.02}_{-0.01}$ \\
\end{tabular}
\end{ruledtabular}
\end{table}

We attempted to reproduce the observed mass spectra
with a Monte-Carlo-type model calculation that includes an in-medium mass modification of
the $\phi$ mesons based on the theoretical predictions of~\cite{HL92, oset,cabrera,KWW}.
We assumed the density dependence of the $\phi$ meson mass as
$m_{\phi}(\rho)/m_{\phi}(0) = 1-k_1(\rho/\rho_{0})$, where $\rho_{0}$ is
the normal nuclear density~\cite{HL92}.
To reproduce the large amount of excess in our data
(22\% for slow $\phi$ in the copper target data),  
it is necessary to introduce a broadening of
the total width of the $\phi$ ($\Gamma_{\phi}^{\rm tot}$),
or at least of the partial width for the \phiee\ decay ($\Gamma_{\phi}^{ee}$).
When no broadening is introduced, the expected rate          
of in-nucleus decay is just 6\% for the $\phi$ mesons produced
in copper nuclei with $\beta \gamma < 1.25$.  
For the density dependence of the total width broadening,
we assumed $\Gamma_{\phi}^{\rm tot}(\rho)/\Gamma_{\phi}^{\rm tot}(0) = 1 + k_2^{\rm tot}(\rho/\rho_{0})$; $k_2^{\rm tot} \ge 0$.
For $\Gamma_{\phi}^{ee}$ we assumed
$\Gamma_{\phi}^{ee}(\rho)/\Gamma_{\phi}^{ee}(0) = 1 + k_2^{ee}(\rho/\rho_{0})$,
and examined following two cases:
(i) the branching ratio $\Gamma_{\phi}^{ee}/\Gamma_{\phi}^{\rm tot}$
remains unchanged in the medium;
$k_2^{ee} = k_2^{\rm tot}$, and
(ii) $\Gamma_{\phi}^{ee}$ doesn't increase in the medium;
$k_2^{ee} = 0$.
It should be noted that
the ratio $N^{in}_{\phi \to ee}$ /$N^{out}_{\phi\to ee}$ increases
with $\Gamma_{\phi}^{tot}$  even in the case (ii).
Here, $N^{in}_{\phi \to ee}$ ($N^{out}_{\phi\to ee}$)
denotes the number of in-medium (in-vacuum) $\phi \to e^+e^-$ decays.
This is because both $N^{in}_{\phi \to ee}$ and $N^{out}_{\phi \to ee}$ decrease but the
latter decreases faster.

We considered that $\phi$ mesons were generated in the target nucleus
according to the nuclear density profile.
This is because we measured the mass-number dependence of the $\phi$ meson production cross section as $\sigma(A) \propto A^1$~\cite{tabaru}.
Generated $\phi$ mesons were traced until the decay point
with the modified pole mass and decay width according to the nuclear density.
The decay probability increases in the medium due to the width broadening.
We used the Woods-Saxon distribution for the nuclear density profile:
$\rho/\rho_0 \propto (1+\exp((r-R)/\tau))^{-1}$, where $R = 4.1 (2.3)\ {\rm fm}$,
and $\tau  = 0.50 (0.57)\ {\rm fm}$ for the copper (carbon) target.

\begin{figure}
\includegraphics[width=8.6cm,height=10.7cm]{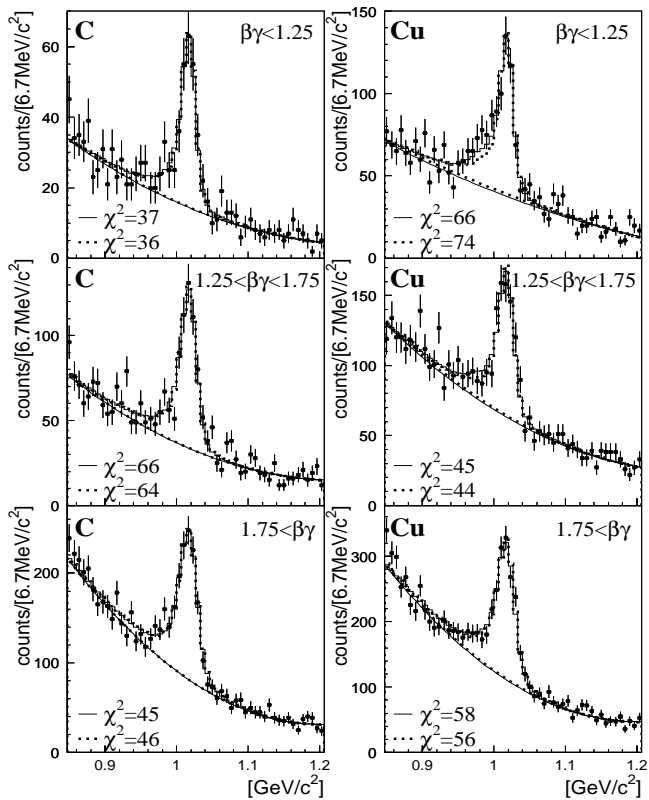}
\caption{
The best fit results in the case (i) (solid line) and (ii) (dotted line).
The shift parameter $k_1$, and broadening parameters
$k_2^{\rm tot}$ and $k_2^{ee}$, are common to the six spectra.
The values of $\chi^2$s are the sum of
the squares of the deviations over 54 data points in each panel.
\label{fig:fitmod}}
\end{figure}

\begin{figure}
\includegraphics[width=8.6cm]{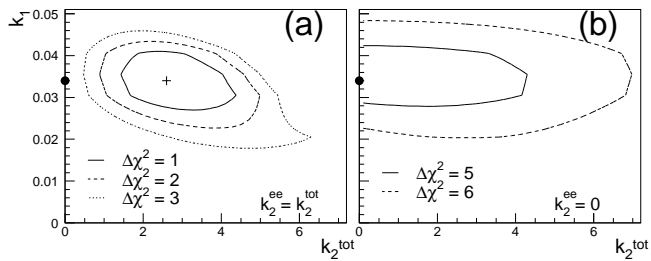}
\caption{ Confidence ellipsoids for the modification parameters
$k_1$ and $k_2^{\rm tot}$
in the case (i) in (a) and (ii) in (b).
The values of $\Delta \chi^2$'s in both panels are
the differences from the $\chi^2_{\rm min} ( = 316.4 )$
at the best fit point in the case (i) which is shown by the cross
in the panel (a).
The best fit point in case (ii) is shown by the closed circle
in the panel (b),
and also in (a) since the ordinates are common to the both cases
in the parameter space.
\label{fig:contour}}
\end{figure}

We modified the resonance shape of the $\phi$ meson with the parameters
$k_1$ and $k_2^{\rm tot}$,
then fitted again the observed mass spectra with the same procedure as before.
For $k_2^{ee}$ we examined the two cases as described above.
The best-fit results in both cases are shown in Fig.~\ref{fig:fitmod}.
The modification parameters $k_1$, $k_2^{\rm tot}$ and $k_2^{ee}$,
are common to the six spectra.
Figure~\ref{fig:contour} shows the confidence ellipsoids for the variation of
$\chi^2$ with $k_1$ and $k_2^{\rm tot}$ in the case (i) and (ii).
In the case (i),
obtained best-fit parameters are
$k_1 = 0.034^{+0.006}_{-0.007}$, $k_2^{\rm tot} = 2.6^{+1.8}_{-1.2}$,
and the minimum $\chi^2$ ($\chi^2_{\rm min}$) is $316.4$.
In the case (ii),
best-fit parameters are $k_1 = 0.033^{+0.011}_{-0.008}$, $k_2^{\rm tot} = 0 ^{+5.6}$,
and the $\chi^2_{\rm min}$ is $320.8$.
In both cases the $\chi^2_{\rm min}$ was obtained
with parameter $k_1 \simeq 0.034$, meaning
the pole mass of the $\phi$ meson decreases by 3.4 \%
at normal nuclear density.
The $\chi^2_{\rm min}$ in the case (ii) (=320.8) is larger than
that in the case (i) (=316.4) by 4.4.
When we fitted only the Cu data in the lowest $\beta \gamma$ region,
where the major discrepancy occurs,
the case (ii) was rejected at 99\% C.L. and 
the best-fit parameters of $k_1$ and $k_2^{\rm tot}$ do not change,
while the case (i) was not rejected.
The data thus favor an increase of $\Gamma_{\phi}^{ee}$ by a factor of
3.6 at normal nuclear density.

In summary, we investigated the mass modification of the $\phi$ mesons
by studying the \ee\ invariant mass distributions obtained in 12 GeV $p+A$ reactions. 
The data obtained with a copper target revealed a significant excess
on the low-mass side of the $\phi$ meson peak
in the $\beta\gamma_\phi < 1.25 $ region.
This observation is consistent with the picture of the $\phi$ modification
in the nucleus.
Thus, we conclude that in addition to our earlier publication on $\rho / \omega$
modification~\cite{ozawa,naruki}, this study experimentally
verified vector meson mass modification at normal nuclear density.

We would like to thank all the staff members of KEK-PS, particularly the
beam channel group for their helpful support. This study was partly funded by the
Japan Society for the Promotion of Science,
RIKEN Special Postdoctoral Researchers Program, and
a Grant-in-Aid for Scientific Research from the Japan Ministry of Education,
Culture, Sports, Science and Technology (MEXT).
Finally, we thank the staff members of
RIKEN Super Combined Cluster (RSCC) and RIKEN-CCJ.

\end{document}